\documentclass[aps,pre,twocolumn,superscriptaddress,showpacs,amsmath,amssymb,footinbib,longbibliography]{revtex4-2}
\usepackage[english]{babel}
\usepackage{amssymb,amsbsy,amsmath}
\usepackage{graphicx}
\usepackage{dcolumn}
\usepackage{bm}
\usepackage{subfigure}
\usepackage{color}
\usepackage{textcomp}
\usepackage[colorlinks,bookmarks=false,citecolor=blue,linkcolor=red,urlcolor=blue]{hyperref}

\newcommand{\op}[1]{\fontdimen12\textfont3=2pt\fontdimen12\scriptfont3=1.4pt\!\null\mathop{\protect\vphantom{#1}\smash{#1}}\limits_{\sim}\null\!}
\newcommand{\xref}[1]{\protect\ref{#1}}
\newcommand{\figref}[1]{Fig.~\protect\ref{#1}}
\newcommand{\fmref}[1]{(\protect\ref{#1})}
\def\bra#1{\langle \, {#1} \, | \,}
\def\ket#1{\, | \, {#1} \, \rangle}
\newcommand{\braket}[2]{\langle \, {#1} \, | \, {#2} \, \rangle}

\renewcommand{\eqref}[1]{Eq.~(\protect\ref{#1})}

\def\erw#1{\langle \, {#1} \, \rangle}

\begin{document}
\title{Non-ergodic one-magnon magnetization dynamics of the antiferromagnetic delta chain}

\author{Florian Johannesmann}
\author{Jannis Eckseler}
\author{Henrik Schl\"uter}
\author{J\"urgen Schnack}
\email{jschnack@uni-bielefeld.de}
\affiliation{Fakult\"at f\"ur Physik, Universit\"at Bielefeld, Postfach 100131, D-33501 Bielefeld, Germany}

\date{\today}

\begin{abstract}
We investigate the one-magnon dynamics of the antiferromagnetic delta chain as a paradigmatic example 
of tunable equilibration. Depending on the ratio of nearest and next-nearest exchange interactions 
the spin system exhibits a flat band in one-magnon space -- in this case equilibration happens only 
partially, whereas it appears to be complete with dispersive bands as generally expected for generic
Hamiltonians. We provide analytical as well as numerical insight into the phenomenon.
\end{abstract}

\keywords{Spin systems, Observables}

\maketitle

\section{Introduction}

Recent theoretical investigations on foundations of thermodynamics focus on 
equilibration as well as thermalization in closed quantum systems under unitary time evolution. 
The road to a deeper understanding was paved by seminal papers of Deutsch, Srednicki and 
many others 
\cite{Deu:PRA91,Sre:PRE94,ScF:NPA96,Tas:PRL98,RDO:N08,PSS:RMP11,ReK:NJP:12,SKN:PRL14,GoE:RPP16,AKP:AP16,BIS:PR16,WDL:PRB17}.
In simple words, the accepted expectation is that generic Hamiltonians, i.e.\ Hamiltonians that are
not special but rather represent a class of similar Hamiltonians, lead to equilibration for the vast majority 
of initial states. In this context it appears interesting to understand 
the \emph{untypical} behavior seen for \emph{special}
Hamiltonians or special states such as quantum scar states \cite{TMA:NP18,TMA:PRB18,HCP:PRL19,MHS:PRB20,KMH:PRB20,PVB:NC21}.

For numerical studies, spin systems are the models of choice both since they are numerically feasible 
due to the finite size of their Hilbert spaces as well as they are experimentally accessible for
instance in standard investigations by means of electron parametric resonance (EPR), 
free induction decay (FID), or in atomic traps, see e.g.\
\cite{CAB:PRB97,MNR:PRB09,CMD:PRL09,ARM:PRL07,JHA:A21,VSS:NJP21}.
In such systems, observables assume expectation values that are practically indistinguishable from 
the prediction of the diagonal ensemble for the vast majority of all late times of their time evolution
under very general and rather not restrictive conditions, see e.g.\ \cite{ReK:NJP:12,SHP:PRE13,BaR:PRL17,SJS:PRB17,SJD:PRE17,RJD:PRB18,RJK:PRB19,BHK:RMP21}.

\begin{figure}[h]
\centering
\includegraphics*[width=0.85\columnwidth]{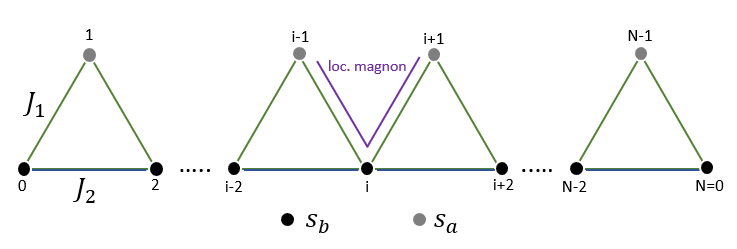}
\includegraphics*[width=0.79\columnwidth]{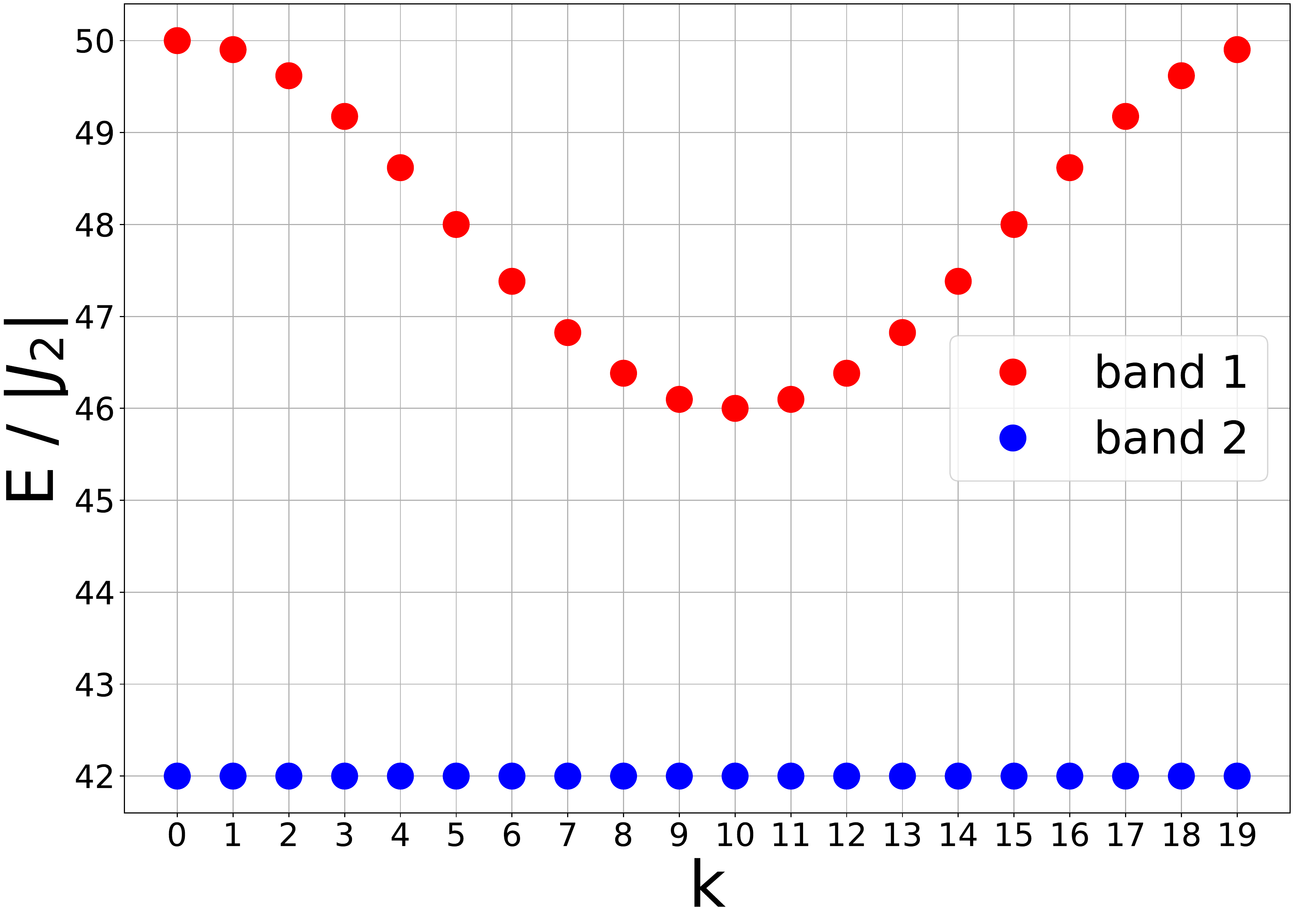}
\caption{Top: Structure of the delta chain with apical spins $s_a$ and basal
spins $s_b$ as well as exchange interactions $J_1$ and $J_2$. 
The spins are numbered $0,1,\dots, N-1$.
An independent localized one-magnon state is highlighted.
Bottom: Energy eigenvalues in one-magnon space for  
$N = 40$, $J_1 = -2$, $ J_2 = -1$ and $s_a = s_b = \frac{1}{2}$.
The momentum quantum number $k$ (wave number) runs from $0$ to $N/2-1$.
}
\label{dispersion}
\end{figure}

In the present paper we investigate the paradigmatic spin delta chain in the Heisenberg model
which becomes special for a certain 
ratio of the two defining exchange interactions $J_1$ and $J_2$,
see \figref{dispersion}. For $J_2/J_1=1/2$ the system exhibits a 
flat band in one-magnon space or equivalently independent localized one-magnon eigenstates of the Hamiltonian,
a phenomenon that has been attracting great attention for more than 20 years now, see e.g.\
\cite{MiT:CMP93,SSR:EPJB01,SHS:PRL02,BlN:EPJB03,RSH:JPCM04,SRM:JPA06,ZhT:PRB04,DRH:PRB10,MHM:PRL12,DRM:IJMP15,LAF:APX18,TDD:ZN20,ROS:PRB22}.
In the context of equilibration, flat bands are interesting since they give rise to zero group velocity
and thus result in a special form of (partial) localization, sometimes also termed 
disorder-free localization \cite{MHS:PRB20}.

Since the one-magnon space of the delta chain hosts only two energy bands (two spins per unit cell)
the quantum problem can be solved analytically. We will present both analytical as well as 
numerical solutions of the time-dependent Schr\"odinger equation and in particular investigate
the magnetization dynamics with and without flat band. We will provide analytical insight into 
which parts of an initial state will not participate in the process of equilibration.
Our results can be qualitatively transferred to other flat-band systems such as kagome, 
square kagome, or pyrochlore spin systems.

The paper is organized as follows. In Section \ref{sec-2} we introduce 
the model, the concept of independent localized magnons as well as the major results.
Section~\xref{sec-3} provides the technical details.
The article closes with a discussion in Section~\ref{sec-4}.


\section{One-magnon dynamics of the delta chain}
\label{sec-2}


The antiferromagnetic delta chain is displayed in \figref{dispersion}~(top).
Assuming periodic boundary conditions, $N\equiv 0$, it is modelled in the Heisenberg model
as
\begin{align}
\label{E-2-0}
\op{H} = -2 J_1 \sum_{i=0}^{N-1} \op{\vec{s}}_i \cdot \op{\vec{s}}_{i+1} - 
2 J_2 \sum_{j=0}^{\frac{N}{2}-1} \op{\vec{s}}_{2j} \cdot \op{\vec{s}}_{2j+2} \ ,
\end{align}
where $\op{\vec{s}}_i$ denote spin vector operators and $J_1<0$ as well as $J_2<0$
are antiferromagnetic exchange interactions.
The model can be treated analytically in one-magnon space, i.e., when the total
magnetic quantum number is given by $M=N(s_a + s_b)/2 - 1$. 
Since the chain hosts two
spins per unit cell the eigenenergies are split into two bands of which 
one is flat for $\alpha=J_2/J_1=1/2$, compare  \figref{dispersion}~(bottom). 
In the later case, one can transform the states 
of the flat band into independent localized one-magnon states,
see \figref{dispersion}~(top) and e.g.\ \cite{SSR:EPJB01,DRM:IJMP15}
\begin{eqnarray}
\label{E-2-1}
\ket{\phi_\mu^0} 
&=& 
\frac{1}{\sqrt{6}}
\left(
\frac{1}{\sqrt{2s_a}}
\op{s}_{\mu-1}^{-}
-
\frac{2}{\sqrt{2s_b}}
\op{s}_\mu^{-}
+
\frac{1}{\sqrt{2s_a}}
\op{s}_{\mu+1}^{-}
\right)
\nonumber
\\
&& 
\times \ket{\Omega}
\ ,
\\
\ket{\Omega}
&=&
\ket{m_0=s_b, m_1=s_a, \dots m_{N-1}=s_a}
\nonumber
\ ,
\end{eqnarray}
where $\mu$ is the position of the basal spin about which the localized magnon is centered,
and $\ket{\Omega}$ denotes the magnon vacuum, i.e., the fully polarized state.
Localized independent one-magnon states have also been 
termed ``compact localized states" recently \cite{CHJ:A22}, we will refer
to them simply as localized states throughout this article.

One may expect that the dynamics is different in the case of a flat band
compared to the generic case of dispersive bands. Qualitatively, 
the argument can be expressed 
in two ways: (1) Since one band is flat, the group velocity of these states 
is strictly zero, and therefore parts of a wave function belonging to the flat band
will not move and therefore never equilibrate or thermalize. (2) Likewise 
one can argue, that the independent localized one-magnon states are stationary 
and contributions of them to a wave function stay localized where they started
initially. Technically, the details are a bit more intricate since the 
localized one-magnon states are not mutually orthogonal; we will
elaborate on this in Sec.~\xref{sec-3}.

The following figures demonstrate the discussed dynamics by showing the local 
magnetization for all sites $i=0, \dots, N-1$, i.e., 
\begin{eqnarray}
\label{E-2-2}
\erw{\op{s}_i^z}_t 
&=& 
\bra{\Psi(t)} \op{s}_i^z \ket{\Psi(t)}
\\
\ket{\Psi(0)} 
&=& \frac{1}{\sqrt{2s_j}}\op{s}_{j}^{-}\ket{\Omega}
\ ,
\end{eqnarray}
starting with a single spin flip at site $j$ at $t=0$. We evaluated the dynamics
both numerically as well as analytically, the latter is shown \cite{Joh:21,Eck:21}.

\begin{figure}[ht!]
\centering
\includegraphics*[width=0.49\columnwidth]{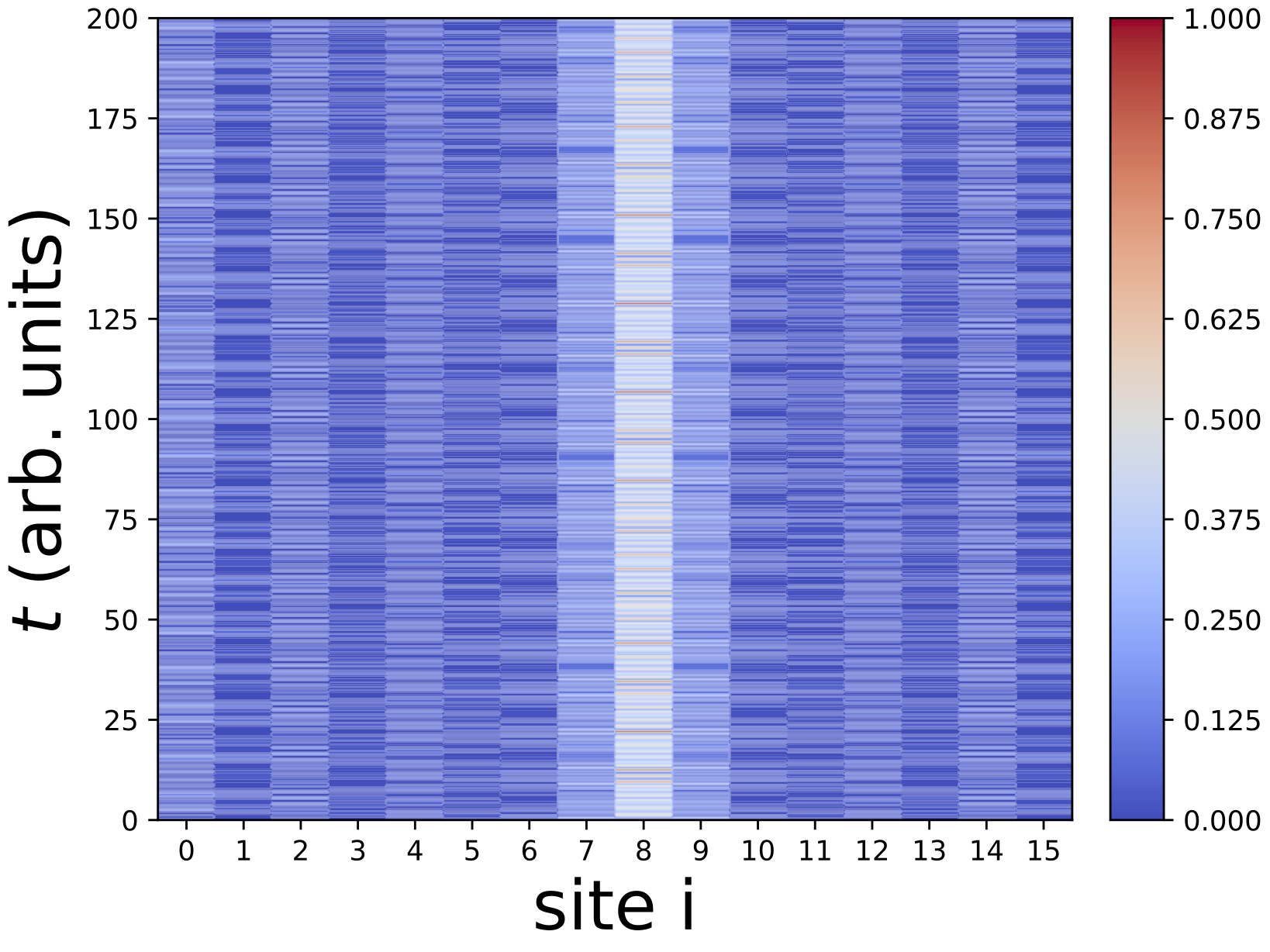}
\includegraphics*[width=0.49\columnwidth]{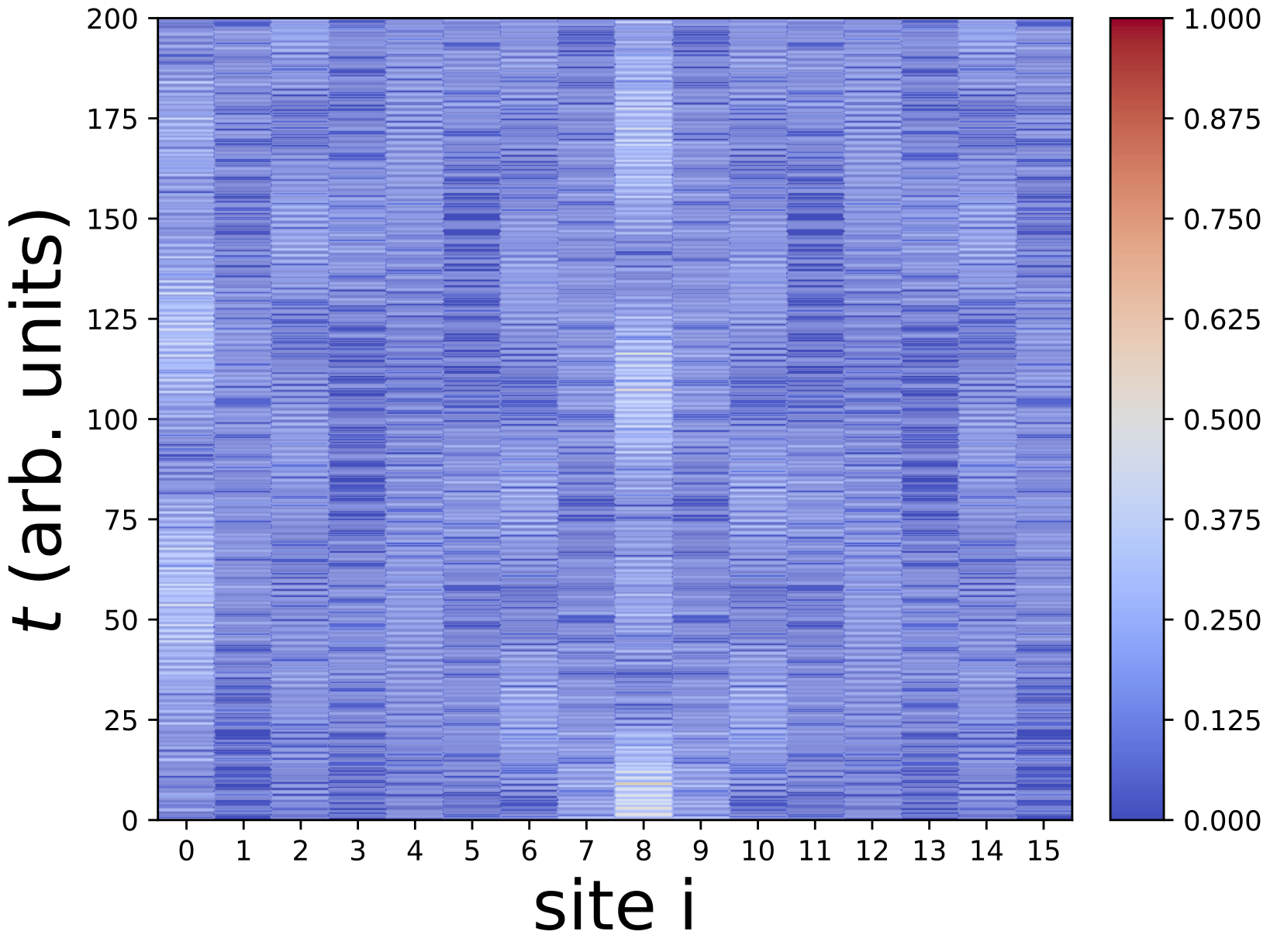}
\caption{$N=16, s_a=s_b=s=1/2, \ket{\Psi(0)} = \frac{1}{\sqrt{2s}}\op{s}_{8}^{-}\ket{\Omega}$:
Magnetization dynamics for $\alpha = 0.5$ (l.h.s.) as well as  
$\alpha = 0.48$ (r.h.s.). The legend shows $0.5-\erw{\op{s}_i^z}_t$.}
\label{dyn-16-8}
\end{figure}

We start our discussion by looking at single spin flips at a basal site $j$. One expects 
that these spin flips differ somewhat from flips at apical sites since they overlap
only with one localized magnon whereas the latter 
overlap with two localized magnons, compare \figref{dispersion}.

Figure~\xref{dyn-16-8} shows the magnetization dynamics for $N=16$ 
and $s_a = s_b = s=\frac{1}{2}$ for the flat-band case $\alpha=J_2/J_1=1/2$
(left) as well as for a nearby Hamiltonian with $\alpha=0.48$ (right), i.e.\ a dispersive
band.
As initial state we choose $\ket{\Psi(0)} = \frac{1}{\sqrt{2s}}\op{s}_{8}^{-}\ket{\Omega}$.
One can see that in the case of a flat band 
a large fraction of the magnetization remains localized 
at the position of the respective independent localized one-magnon state
to which the site of the excitation belongs (sites 7,8,9 in the example)
whereas for the (only slightly) dispersive band the magnetization delocalizes 
across the system. Since the system is rather small one observes to a small extend 
waves that run around the system due to periodic boundary conditions; 
they give rise to interferences.

\begin{figure}[ht!]
\centering
\includegraphics*[width=0.49\columnwidth]{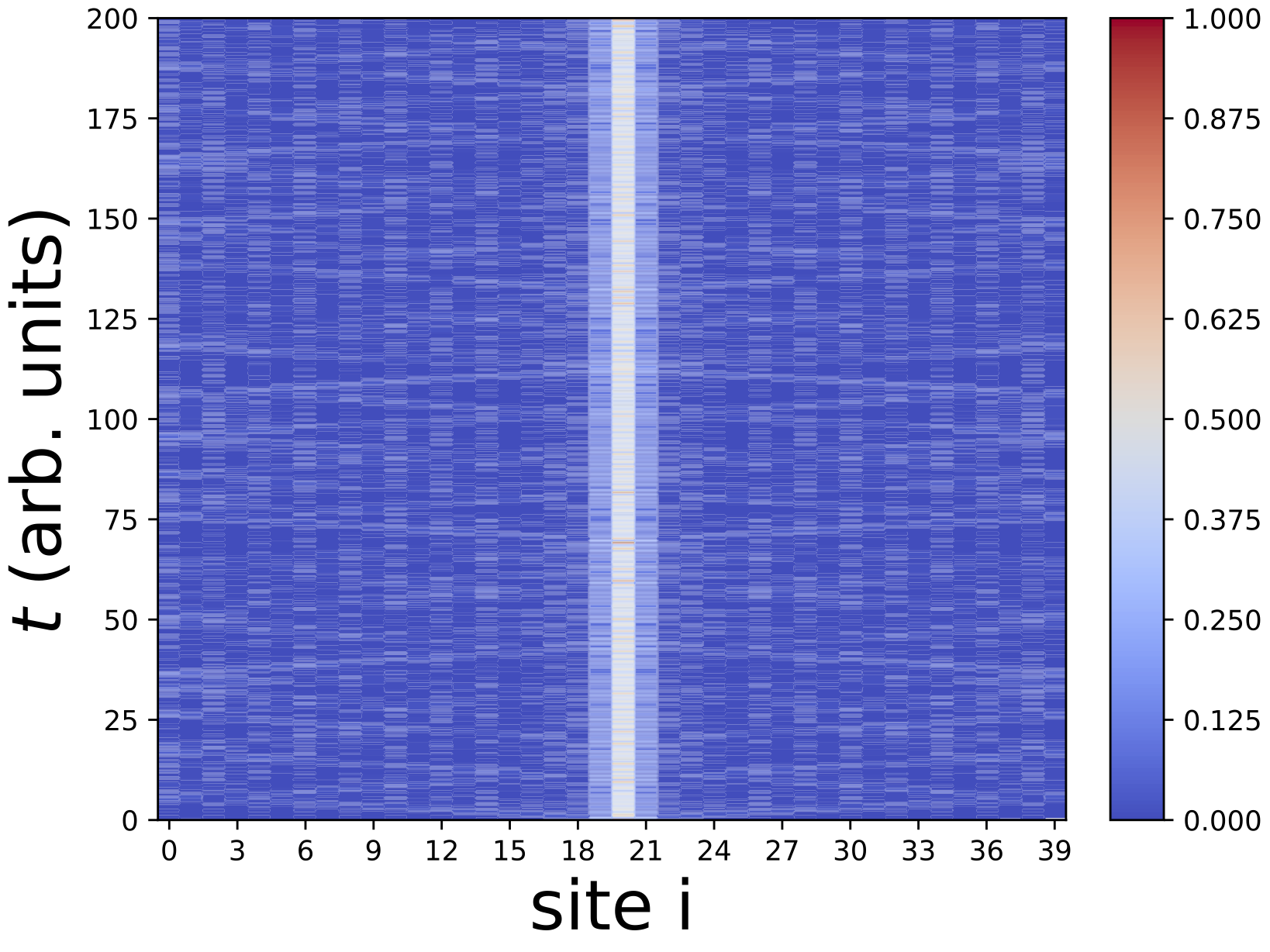}
\includegraphics*[width=0.49\columnwidth]{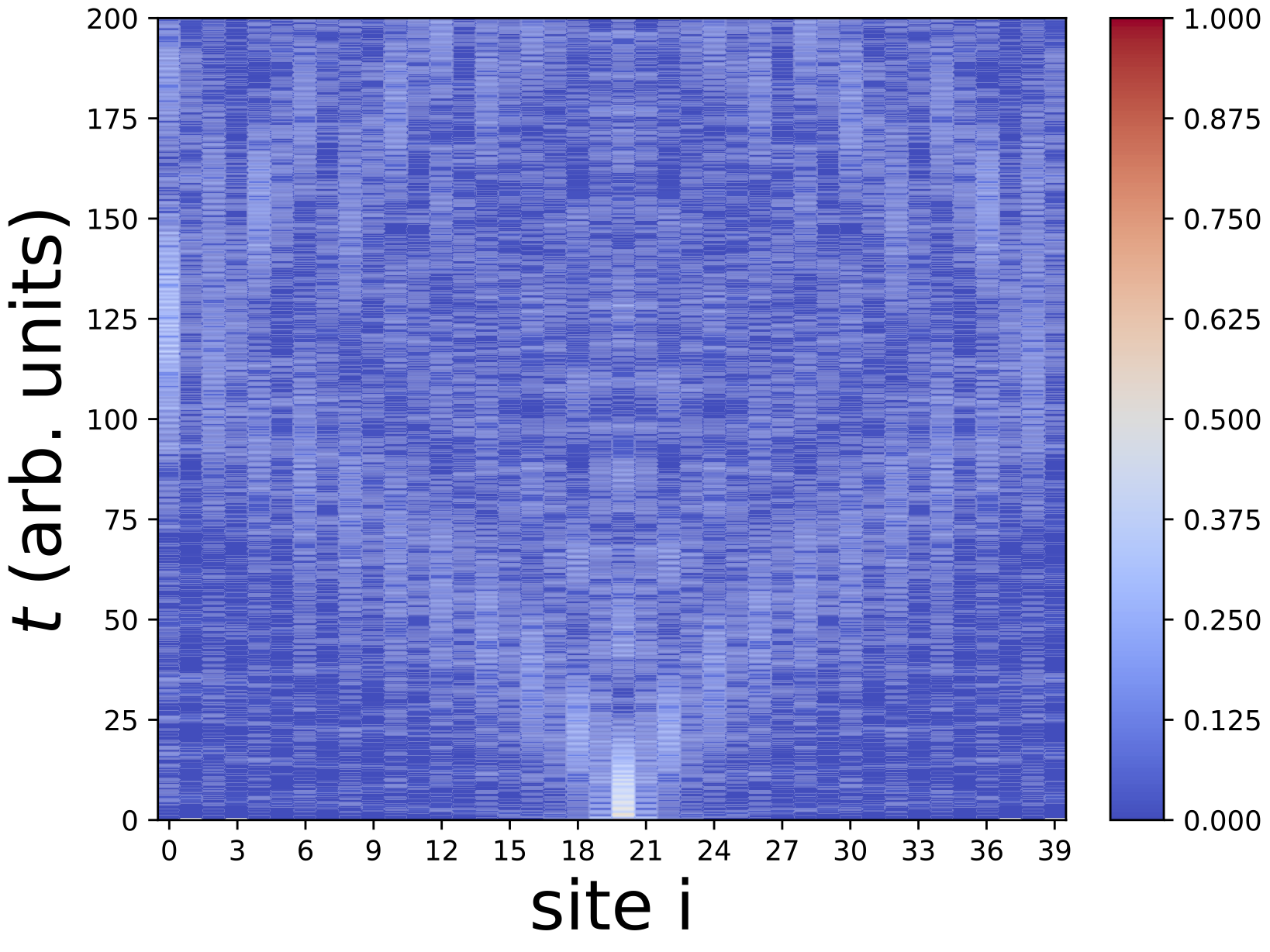}
\caption{$N=40, s_a=s_b=s=1/2, \ket{\Psi(0)} = \frac{1}{\sqrt{2s}}\op{s}_{20}^{-}\ket{\Omega}$:
Magnetization dynamics for $\alpha = 0.5$ (l.h.s.) as well as  
$\alpha = 0.48$ (r.h.s.). The legend shows $0.5-\erw{\op{s}_i^z}_t$.}
\label{dyn-40-20}
\end{figure}

\begin{figure}[ht!]
\centering
\includegraphics*[width=0.49\columnwidth]{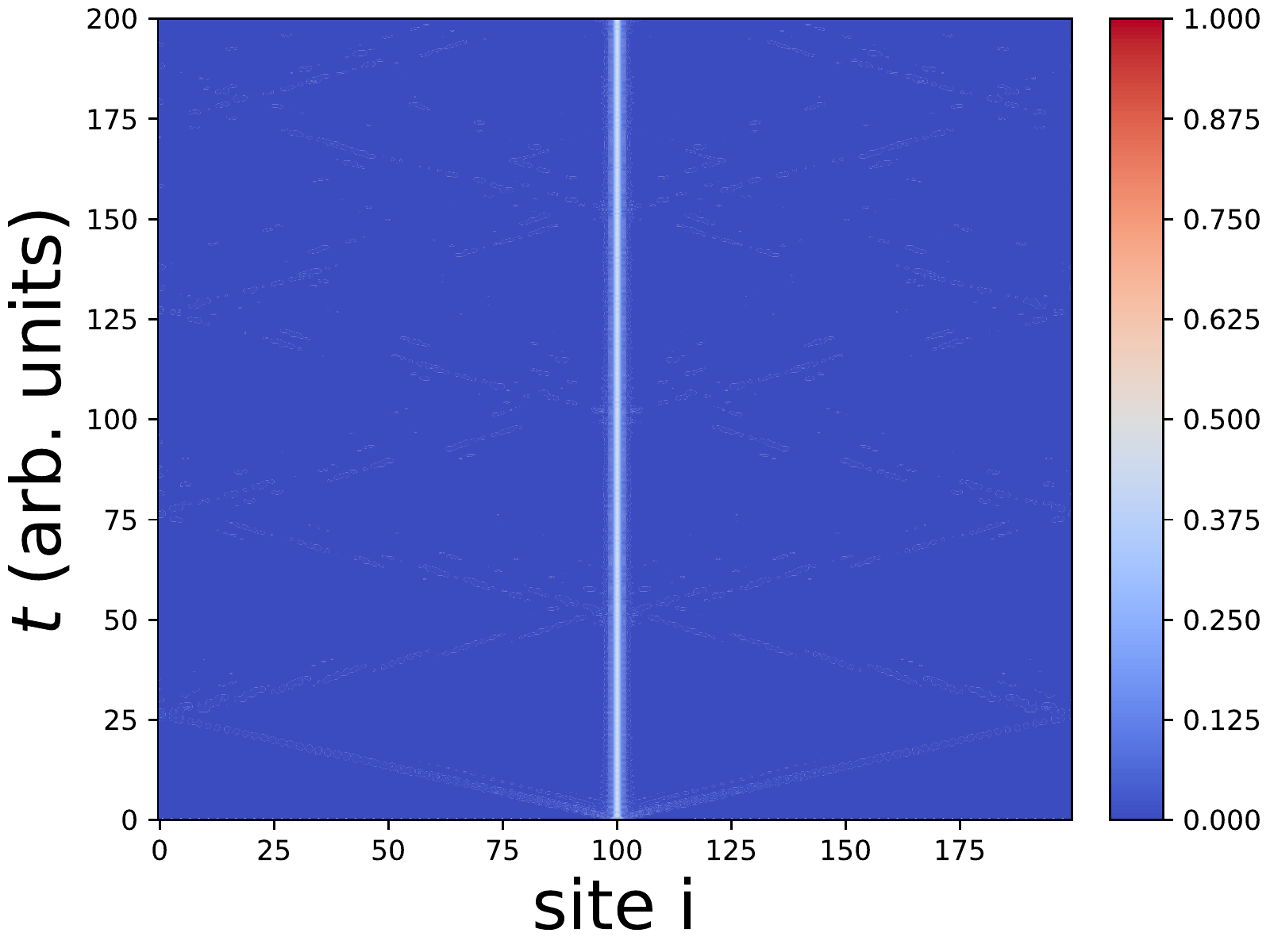}
\includegraphics*[width=0.49\columnwidth]{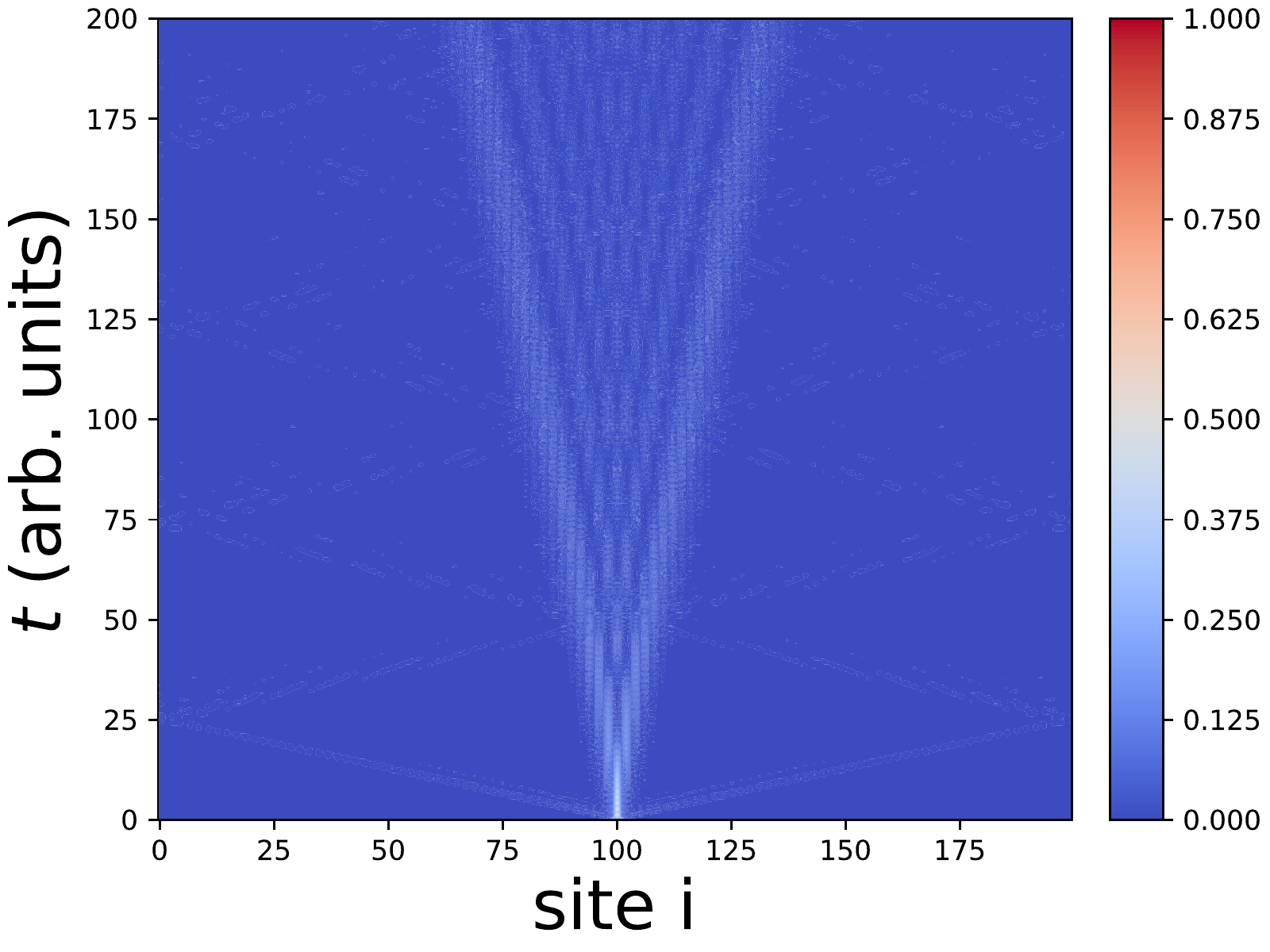}
\caption{$N=200, s_a=s_b=s=1/2, \ket{\Psi(0)} = \frac{1}{\sqrt{2s}}\op{s}_{100}^{-}\ket{\Omega}$:
Magnetization dynamics for $\alpha = 0.5$ (l.h.s.) as well as  
$\alpha = 0.48$ (r.h.s.). The legend shows $0.5-\erw{\op{s}_i^z}_t$.}
\label{dyn-200-100}
\end{figure}

The question is how larger systems behave. To this end we show results for
$N=40$ in \figref{dyn-40-20} as well as $N=200$ in \figref{dyn-200-100}.  
One clearly sees -- left hand sides of both figures --
that a remanent magnetization persists at the site of the 
localized magnon overlapping with the single-spin excitation for the case
of a flat band. In case of dispersive bands the initially maximally localized 
magnetization fluctuation redistributes over the entire system 
(right hand sides of the figures).

\begin{figure}[h]
\centering
\includegraphics*[width=0.49\columnwidth]{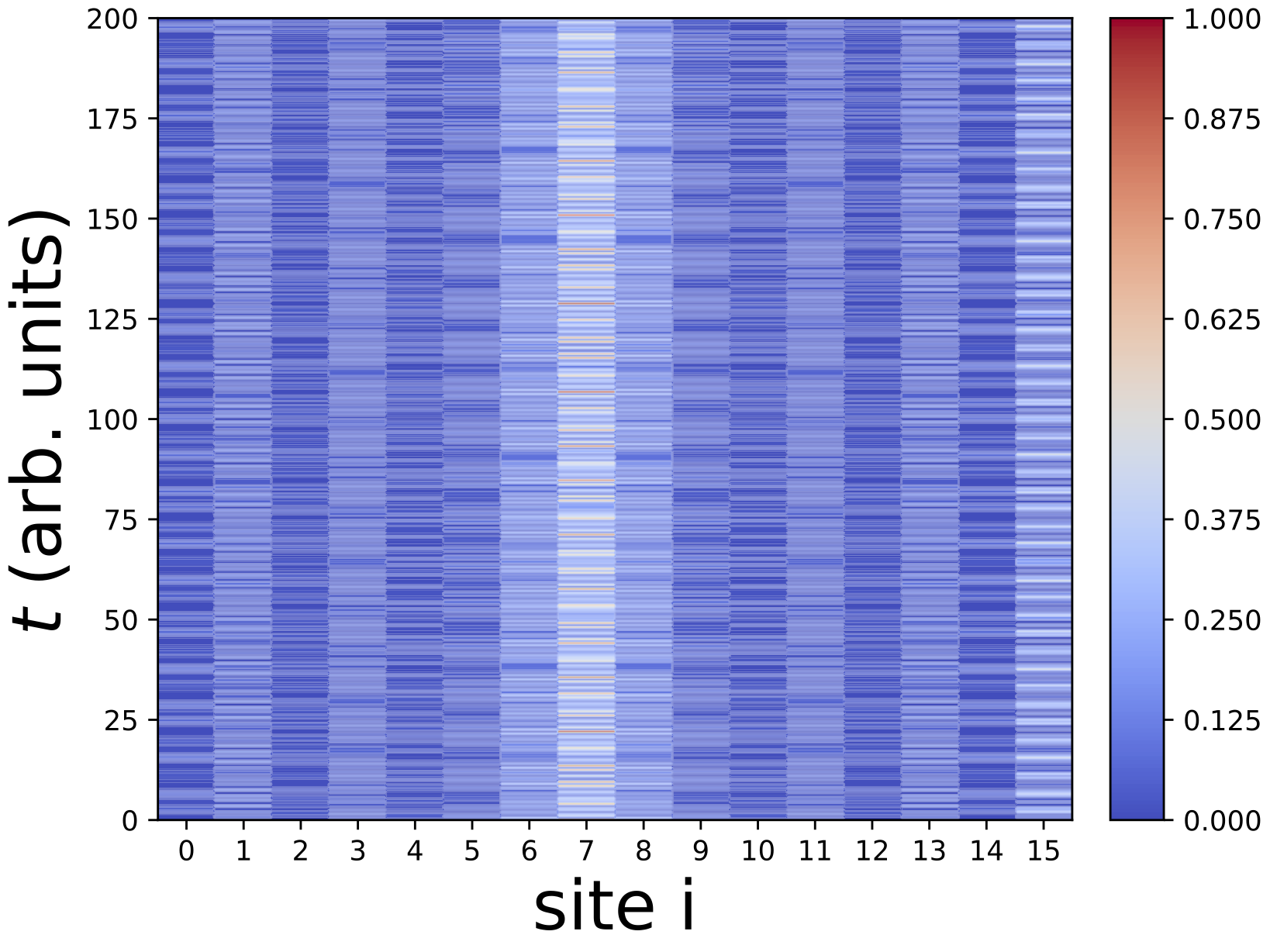}
\includegraphics*[width=0.49\columnwidth]{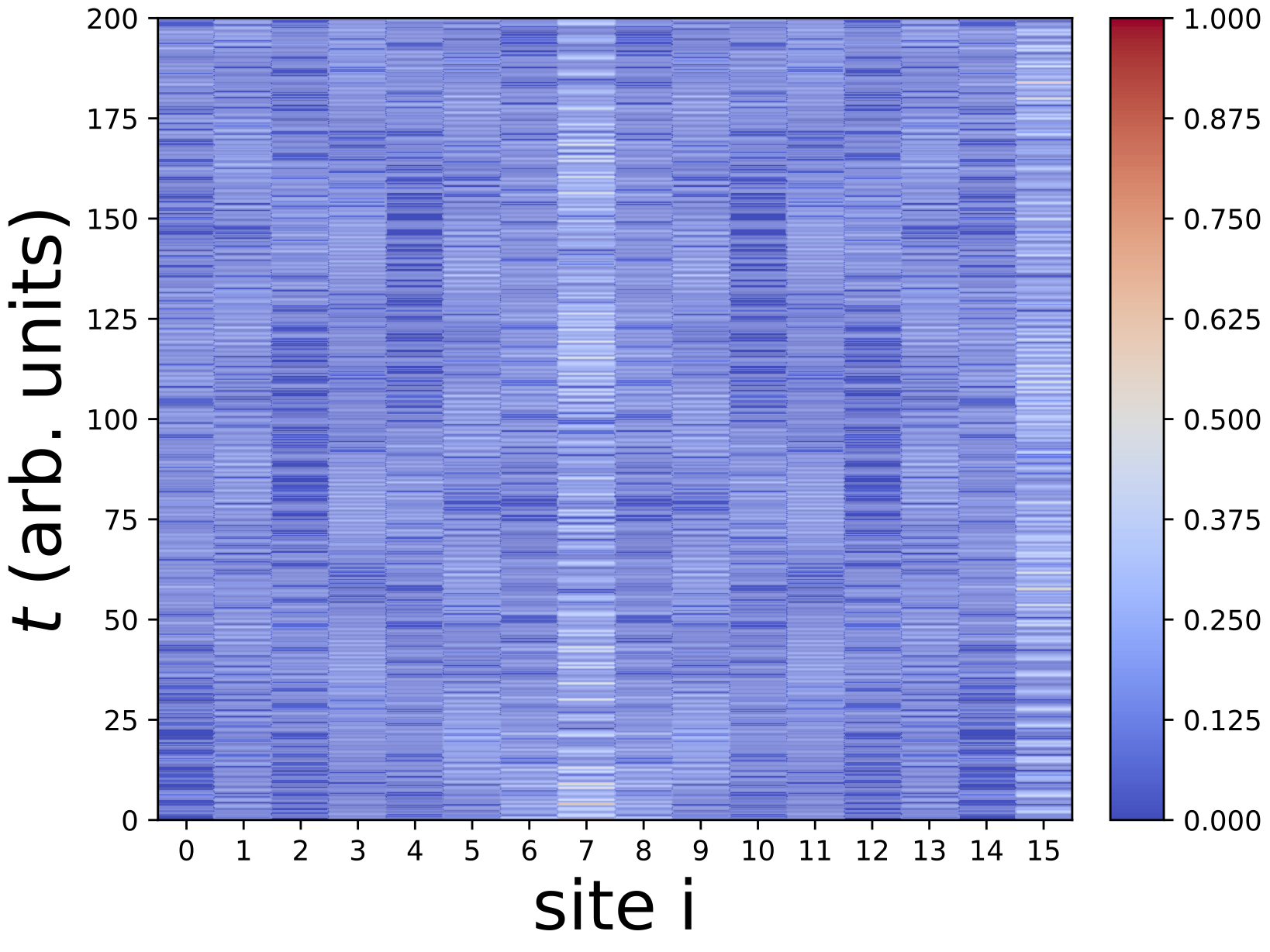}
\caption{$N=16, s_a=s_b=1/2, \ket{\Psi(0)} = \frac{1}{\sqrt{2s_7}}\op{s}_{7}^{-}\ket{\Omega}$:
Magnetization dynamics for $\alpha = 0.5$ (l.h.s.) as well as  
$\alpha = 0.48$ (r.h.s.). The legend shows $0.5-\erw{\op{s}_i^z}_t$.}
\label{dyn-16-7}
\end{figure}

\begin{figure}[h]
\centering
\includegraphics*[width=0.49\columnwidth]{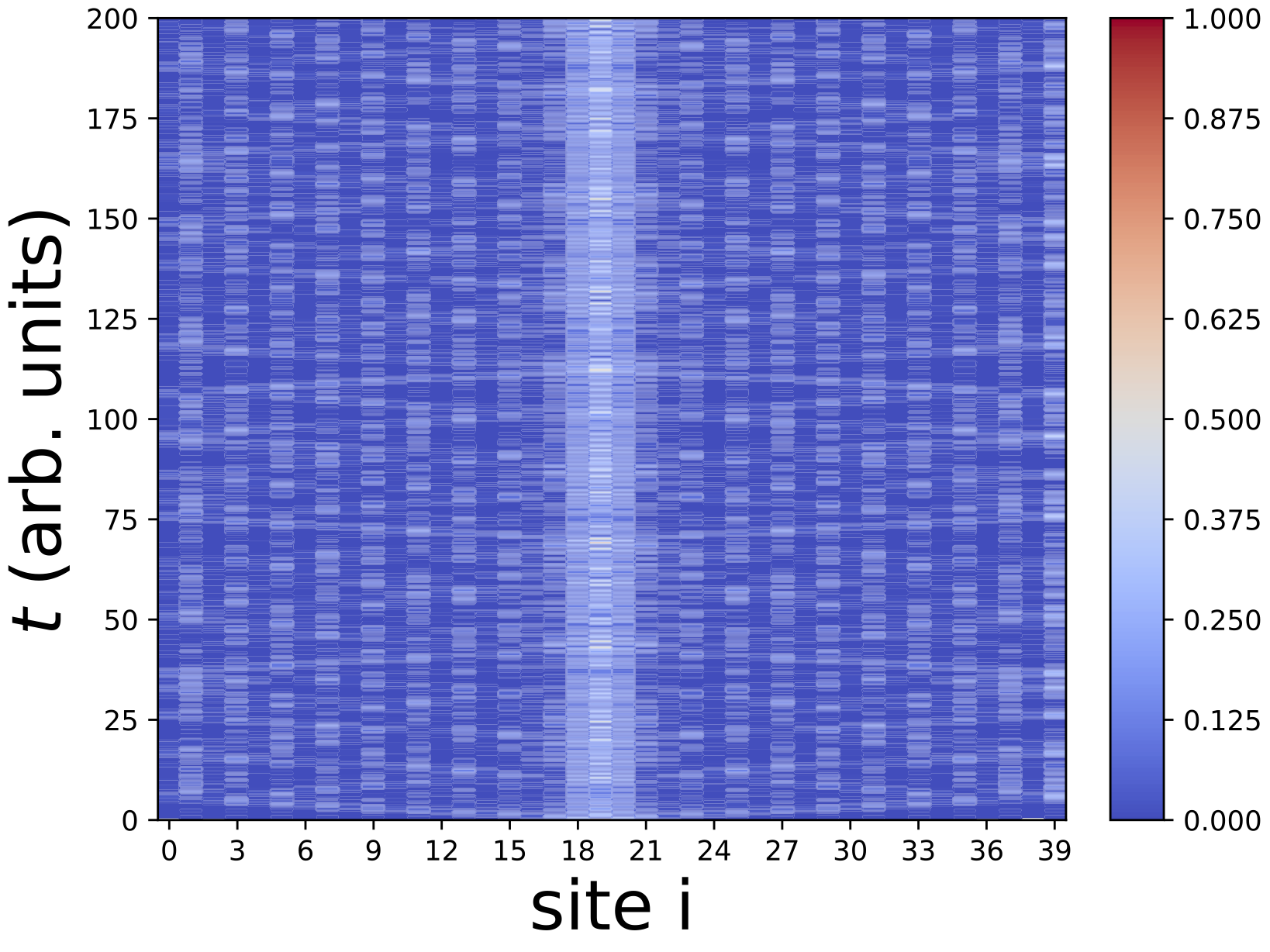}
\includegraphics*[width=0.49\columnwidth]{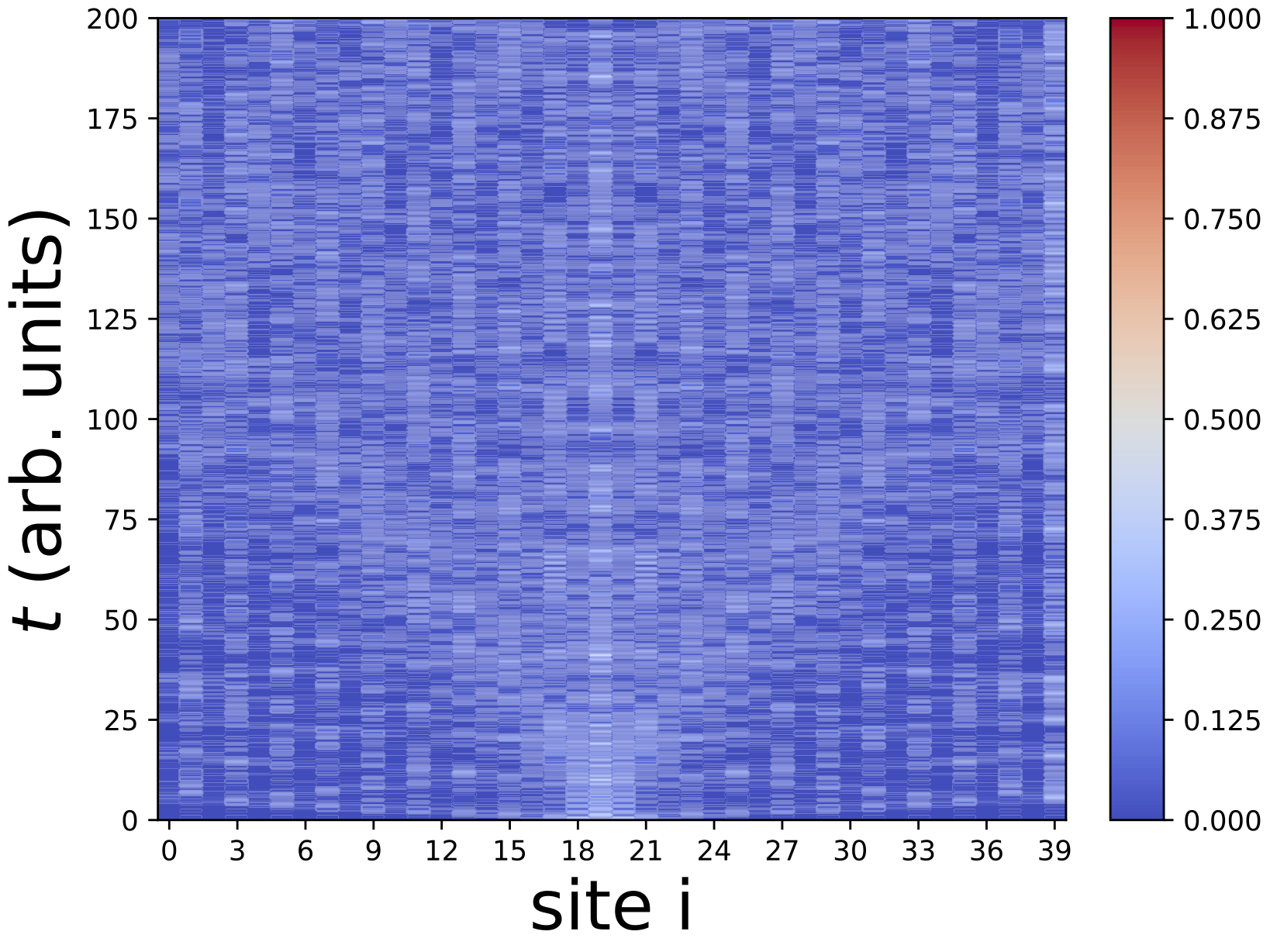}
\caption{$N=40, s_a=s_b=s=1/2, \ket{\Psi(0)} = \frac{1}{\sqrt{2s}}\op{s}_{19}^{-}\ket{\Omega}$:
Magnetization dynamics for $\alpha = 0.5$ (l.h.s.) as well as  
$\alpha = 0.48$ (r.h.s.). The legend shows $0.5-\erw{\op{s}_i^z}_t$.}
\label{dyn-40-19}
\end{figure}

\begin{figure}[ht!]
\centering
\includegraphics*[width=0.49\columnwidth]{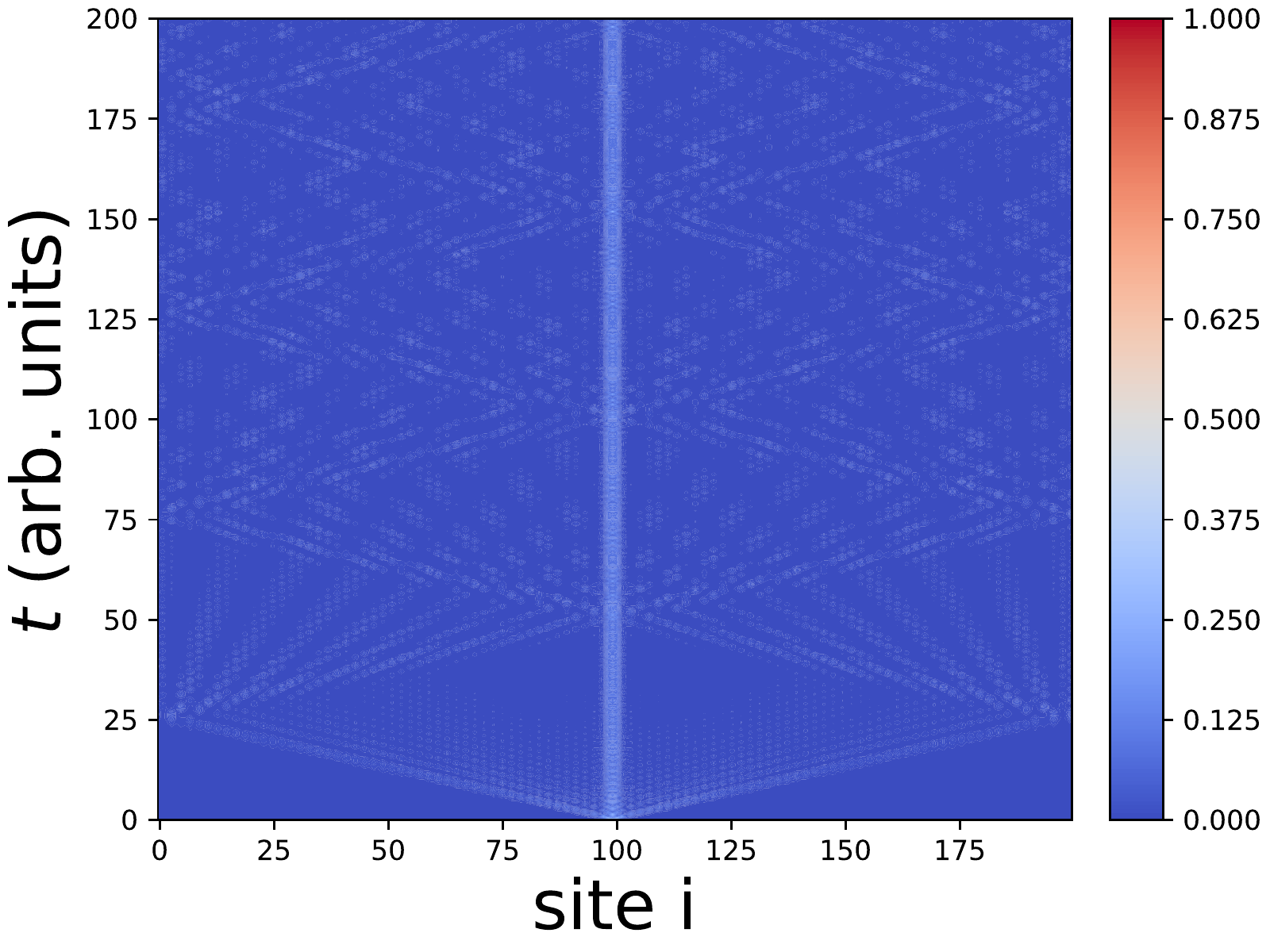}
\includegraphics*[width=0.49\columnwidth]{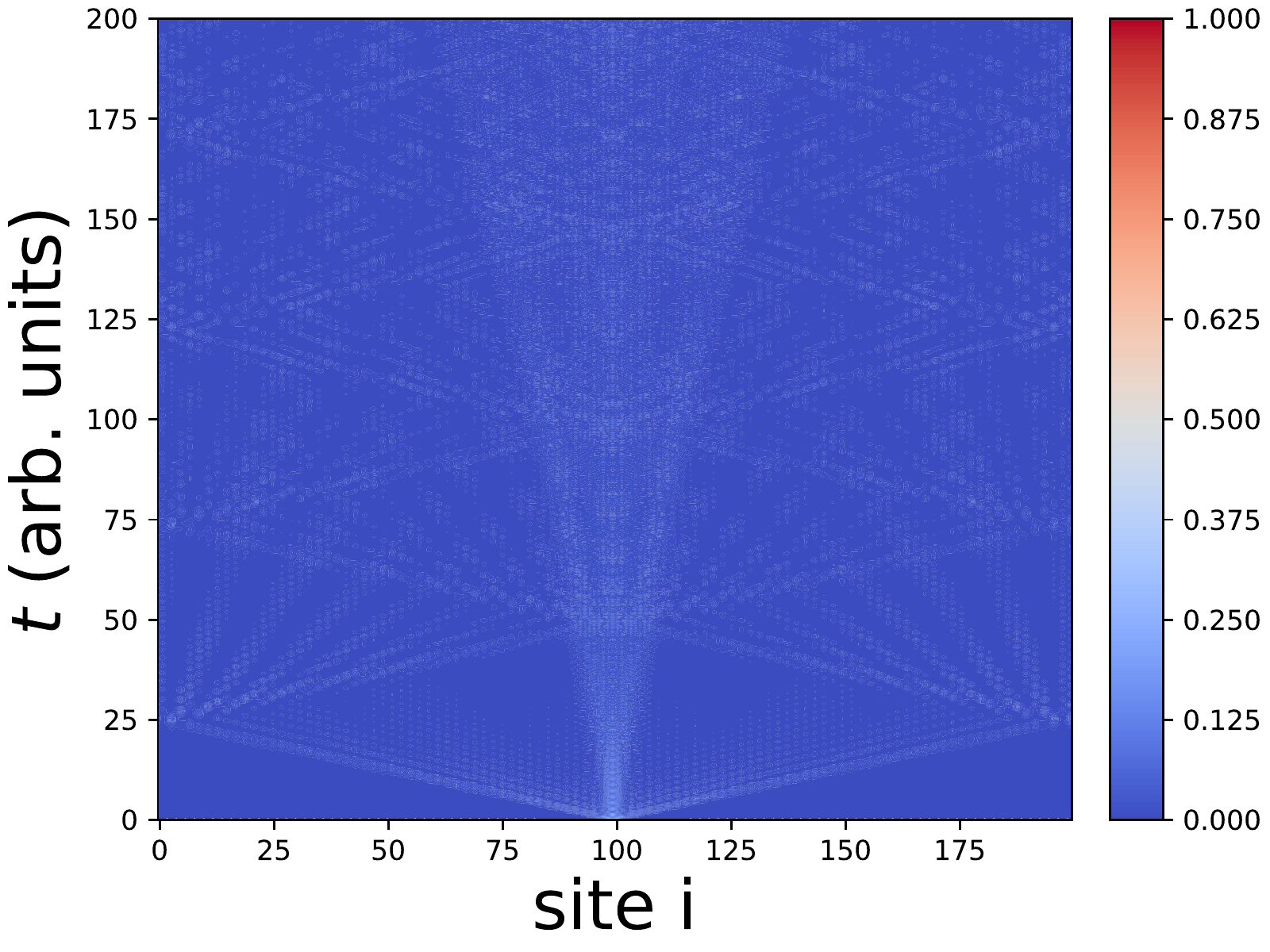}
\caption{$N=200, s_a=s_b=s=1/2, \ket{\Psi(0)} = \frac{1}{\sqrt{2s}}\op{s}_{99}^{-}\ket{\Omega}$:
Magnetization dynamics for $\alpha = 0.5$ (l.h.s.) as well as  
$\alpha = 0.48$ (r.h.s.). The legend shows $0.5-\erw{\op{s}_i^z}_t$.}
\label{dyn-200-99}
\end{figure}

The situation changes somewhat if the spin flip is executed at an
apical site. Such a site belongs to two independent localized 
one-magnon states, therefore the magnetization remains dominantly localized 
across both states. It should also be somewhat smaller since it is now
distributed over 5 sites.

The figures \xref{dyn-16-7}, \xref{dyn-40-19}, and \xref{dyn-200-99}
display the cases of $N=16$, $N=40$, and $N=200$, respectively.
Again the main insight we gain is that for the flat band cases there
is a remanent magnetization distributed about the site of the 
single-spin flip whereas for the (only slightly) dispersive band 
the magnetization redistributes over the entire system.

\begin{figure}[ht!]
\centering
\includegraphics*[width=0.455\columnwidth]{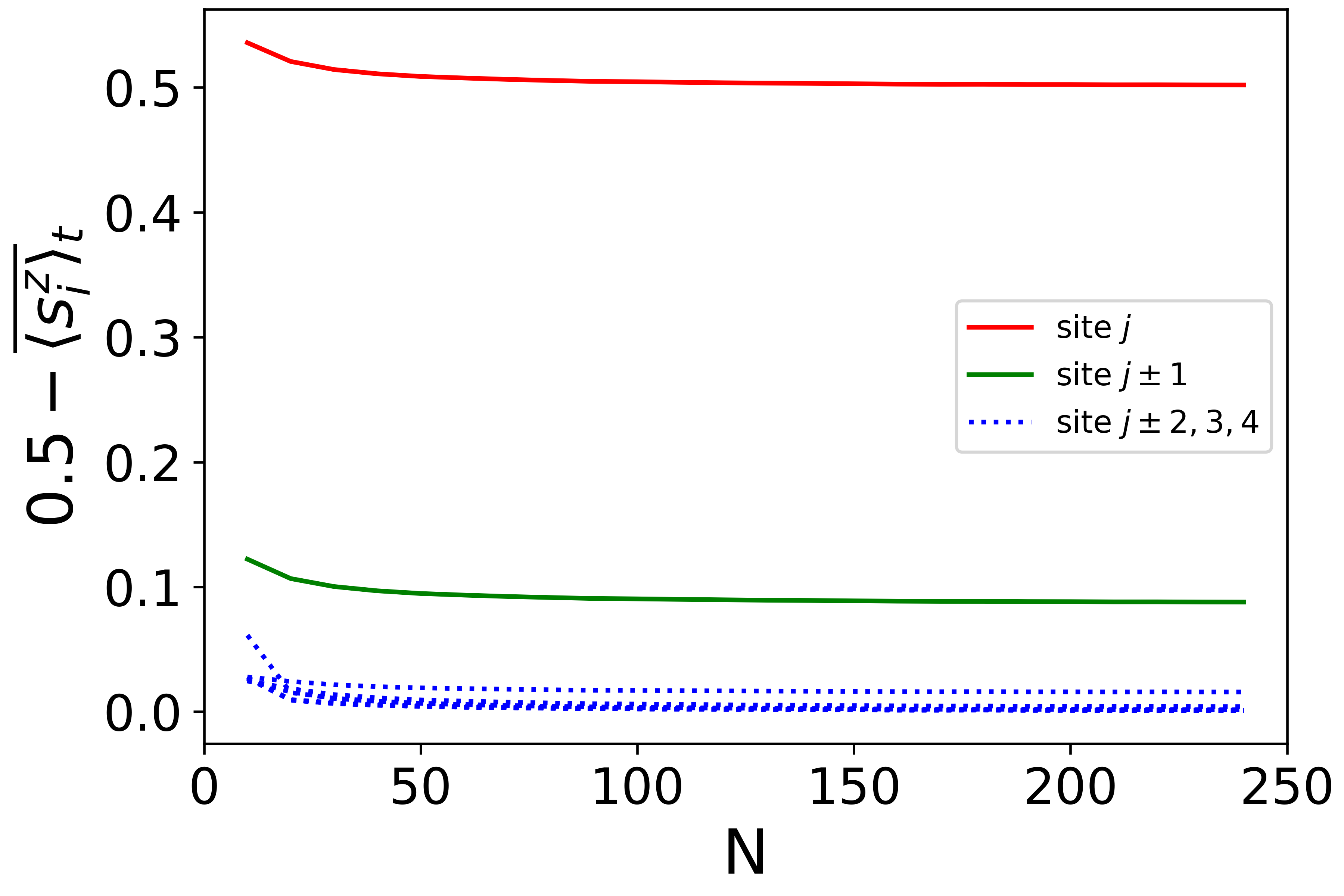}
\includegraphics*[width=0.455\columnwidth]{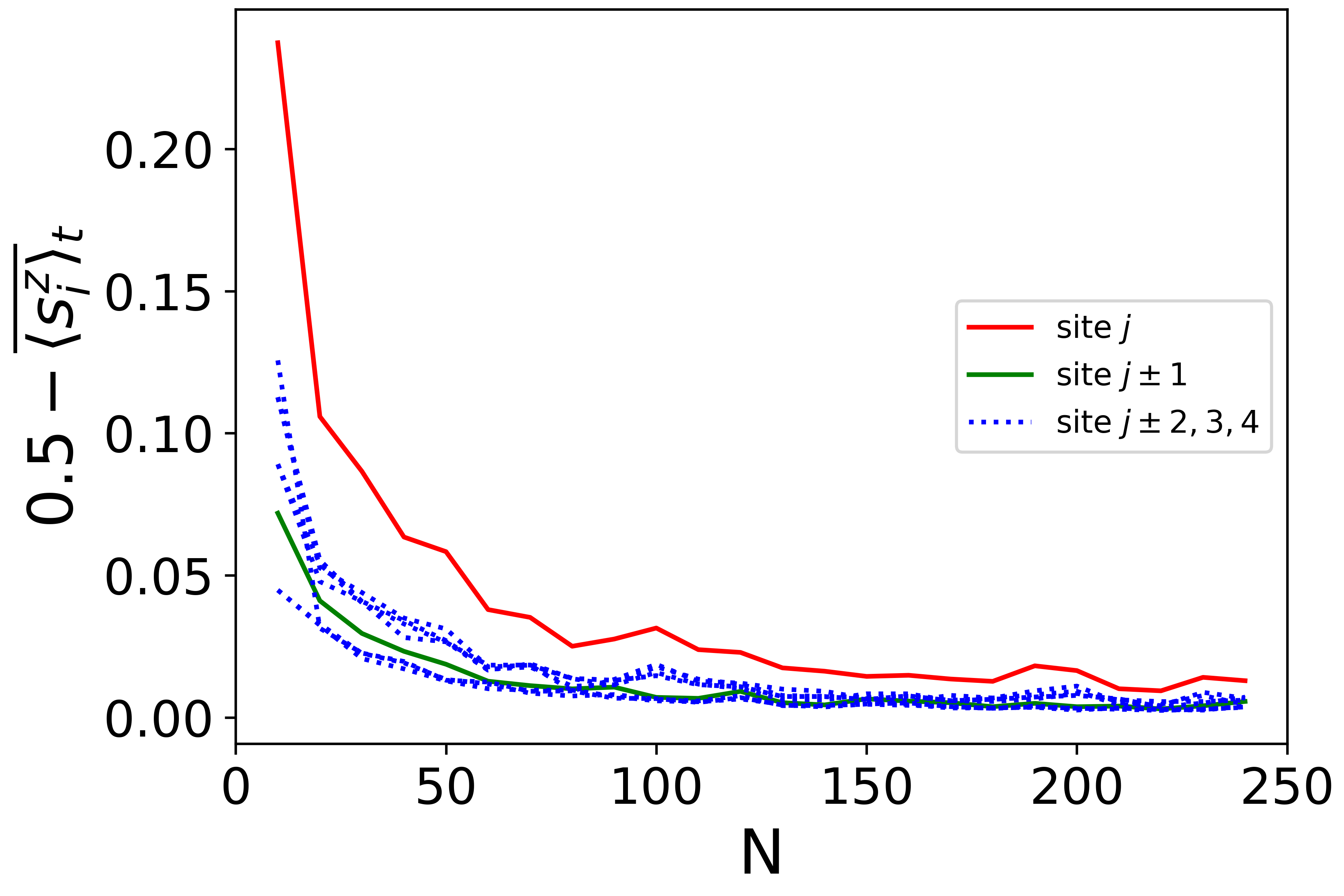}
\caption{Time-averaged local magnetization (above background of magnon vacuum)
$(0.5-\overline{\erw{\op{s}_i^z}}_t)$
at sufficiently late times, 
compare \fmref{E-2-3},
for various sizes of the spin system: $\alpha = 0.5$ (l.h.s.) and
$\alpha = 0.48$ (r.h.s.). All systems were time-evolved over $t=1,000,000$ of 
our time units and then averaged over additional $n_t \Delta t=2,000$ 
time units, compare \fmref{E-2-3}.}\label{dyn-late-times}
\end{figure}

We summarize our results graphically in \figref{dyn-late-times} where we plot
the time-averaged local magnetization, \eqref{E-2-2}, at sufficiently 
late times for various sizes of the spin system, i.e.
\begin{eqnarray}
\label{E-2-3}
\overline{\erw{\op{s}_i^z}}_t 
&=& 
\frac{1}{n_t \Delta t}
\sum_{n=1}^{n_t}
\bra{\Psi(t+n\Delta t)} \op{s}_i^z \ket{\Psi(t+n\Delta t)}
\ .
\end{eqnarray}
We restrict ourselves to single-spin flips at a basal site. 
As one can see on the l.h.s.\ of \figref{dyn-late-times} 
the local magnetization at the
site of the flip drops from one to 1/2 above background. 
At the neighboring sites that belong
to the localized magnon $\ket{\mu=j}$ the local magnetization approaches 
roughly $0.1$ above background. 
This means that out of the initial magnetization fluctuation
about seventy percent remain localized at the respective localized magnon,
a substantial fraction that never equilibrates. The precise contributions
of a local spin flip, that do not perticipate in an equlibrating dynamics 
will be exactly evaluated in Sec.~\xref{sec-3}.

For the case of a dispersive band, shown on the r.h.s. of \figref{dyn-late-times}
one immediately realizes that the magnetization fluctuation due to the single 
spin flip is practically evenly redistributed over the entire system. 
All late-time single-spin expectation values approach the background value of
the magnon vacuum (set to zero) plus $1/N$ for the redistributed single-spin flip.

Finally, since this is not the focus of the paper at hand,
we refer readers interested in the question how the system approaches its long-time 
limit, i.e.\ ballisticly or diffusively to the existing extensive wealth of papers on
that topic \cite{HHC:PRB02,HHC:PRB03,SHP:PRE13,SJS:PRB17,SJD:PRE17,RJD:PRB18,RJK:PRB19,BHK:RMP21}.

\section{Analytical solution for the delta chain}
\label{sec-3}

\begin{figure}[h]
\centering
\includegraphics*[width=0.85\columnwidth]{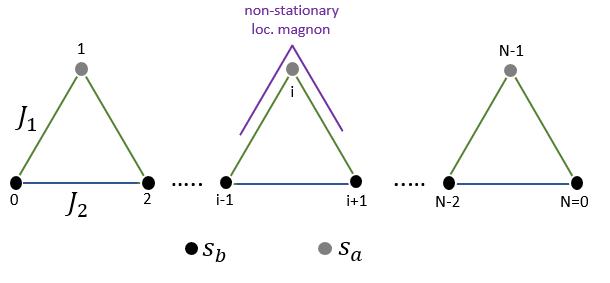}
\caption{Structure of the delta chain with apical spins $s_a$ and basal
spins $s_b$ as well as exchange interactions $J_1$ and $J_2$. 
The spin are numbered $0,1,\dots, N-1$.
A localized non-stationary localized one-magnon state is highlighted.}
\label{loc-non-stat-magnon}
\end{figure}

All results discussed in Sec.~\xref{sec-2} can be obtained either numerically or even analytically.
An analytical solution for the delta chain can be achieved using the symmetries
of the Hamiltonian. One-magnon space is spanned by $N$ states 
$\frac{1}{\sqrt{2s_i}}\op{s}_{i}^{-}\ket{\Omega}$, and thus has got a dimension of $N$.
A unit cell of the chain hosts two spins, one $s_a$ and one $s_b$ spin, respectively,
with a translational symmetry 
\begin{eqnarray}
\label{E-3-1}
\op{T} \ket{m_0, m_1, \dots, m_{N-1}} = \ket{m_{N-2}, m_{N-1}, m_0, \dots}
\ .
\end{eqnarray}
This leads to two bands of energy eigenvalues, each with $N/2$ states with momentum quantum numbers 
$k=0, 1, \dots, N/2$, compare \figref{dispersion}~(bottom). The energy eigenvalues 
$\epsilon_{k, \tau=\pm 1}$ as well as eigenstates $\ket{\epsilon_{k, \tau=\pm 1}}$
can be obtained analytically since the Hamiltonian matrix is only of size $2\times 2$ for each value
of $k$,

\onecolumngrid
\begin{eqnarray}
\label{E-3-2}
\epsilon_{k, 1/2} 
&=& -2 J_1 (N s_a s_b - s_a - s_b) - 2 J_2\left(\frac{N}{2} s_b^2 - s_b\left(1-\cos\left(\frac{4 \pi k}{N}\right)\right)\right)
\\
&&\pm \Bigg\{J_1^2 s_a^2 + 2 J_1 J_2 s_a s_b + (J_1-J_2)^2 + s_b \cos\left(\frac{4 \pi k}{N}\right)
\left(
2 (J_1 - J_2) (J_1 s_a + J_2 s_b) + J_2^2 s_b \cos\left(\frac{4 \pi k}{N}\right)
\right)\Bigg\}^{1/2}
\nonumber
\ ,
\end{eqnarray}
\twocolumngrid

where $\epsilon_{k, 1}$ corresponds to the ``$+$"-sign and $\epsilon_{k, 2}$ to the ``$-$"-sign,
respectively.
For $J_2 = J$ and $J_1 = 2 J$ one obtains
\begin{eqnarray}
\label{E-3-3a}
\epsilon_{k, 1} 
&=& -J s_b \left\{ N (4 s_a + s_b) - 4 \left(1-\cos\left(\frac{4 \pi k}{N}\right)\right)\right\}
\\
\label{E-3-3b}
\epsilon_{k, 2} 
&=& -J \left\{ 4 s_a (N s_b - 2) +s_b (N s_b-8)\right\}
\ ,
\end{eqnarray}
where $\epsilon_{k, 2}$ constitutes the flat band. The local magnetization at site $j$
as displayed in Figs.~\xref{dyn-16-8}-\xref{dyn-200-99}
can analytically be evaluated as \cite{Joh:21}
\begin{eqnarray}
\label{E-3-4}
\bra{\psi(t)}\op{s}_j^z\ket{\psi(t)} 
&&=
\bra{\psi(0)} e^{\frac{i}{\hbar}\op{H} \cdot t} \op{s}_j^z e^{- \frac{i}{\hbar} \op{H} \cdot t} \ket{\psi(0)} 
\\
=\sum_{k, \tau} \sum_{k^{\prime}, \tau^{\prime}} 
&&
\braket{\psi(0)}{\epsilon_{k, \tau}} \bra{\epsilon_{k, \tau}}\op{s}_j^z\ket{\epsilon_{k^{\prime}, \tau^{\prime}}} 
\nonumber
\\
&&
\times\braket{\epsilon_{k^{\prime}, \tau^{\prime}}}{\psi(0)} 
e^{\frac{i}{\hbar} \left( \epsilon_k^{\tau} - \epsilon_{k^{\prime}}^{\tau^{\prime}} \right) \cdot t} 
\nonumber
\ .
\end{eqnarray}
A deeper insight of the magnetization dynamics can be obtained by using a new
basis in one-magnon space that consists of the localized magnons introduced in
\eqref{E-2-1} and \figref{dispersion} complemented by analogous states constructed 
from the upper band,
\begin{eqnarray}
\label{E-3-5}
\ket{\phi_\mu^1} 
&=& 
\frac{1}{\sqrt{6}}
\left(
\frac{1}{\sqrt{2s_b}}
\op{s}_{\mu-1}^{-}
+
\frac{2}{\sqrt{2s_a}}
\op{s}_\mu^{-}
+
\frac{1}{\sqrt{2s_b}}
\op{s}_{\mu+1}^{-}
\right)
\nonumber
\\
&& 
\times \ket{\Omega}
\ .
\end{eqnarray}
We term the latter non-stationary localized magnon states; 
they are depicted in \figref{loc-non-stat-magnon}.
For these states we find $\braket{\phi_\mu^0}{\phi_\nu^1}=0$, but otherwise
they are not orthogonal. Although this complicates their use for
easy (hand-waving) interpretations of the results a little bit, the typical arguments 
we used in Sec.~\xref{sec-2}
resting e.g.\ on overlaps are dominantly correct, i.e.\ up to small technical 
corrections.

A technically correct decomposition of the initial spin flip
at site $j$
\begin{eqnarray}
\label{E-3-6}
\frac{1}{\sqrt{2s_j}}
\op{s}_j^{-}
\ket{\Omega}
=
&&
\sum_{\mu=0, 2, 4, \dots}
c_\mu^{(j)}
\ket{\phi_\mu^0} 
\\
&+&
\sum_{\mu=1, 3, 5, \dots}
c_\mu^{(j)}
\ket{\phi_\mu^1}
\nonumber
\end{eqnarray}
has to be performed e.g.\ by a Householder QR-decomposition.
The coefficients $c_\mu^{(j)}$ are not given by dot products 
(overlaps) between the spin-flip state and the basis states 
as would be the case for an orthonormal basis. However, the 
easy (handwaving) interpretation used in Sec.~\xref{sec-2}
that the spin-flip state has got an overlap with a localized
magnon (or two) and thus remains partially trapped at the site
of the localized magnon remains true. 

\begin{figure}[h]
\centering
\includegraphics*[width=0.85\columnwidth]{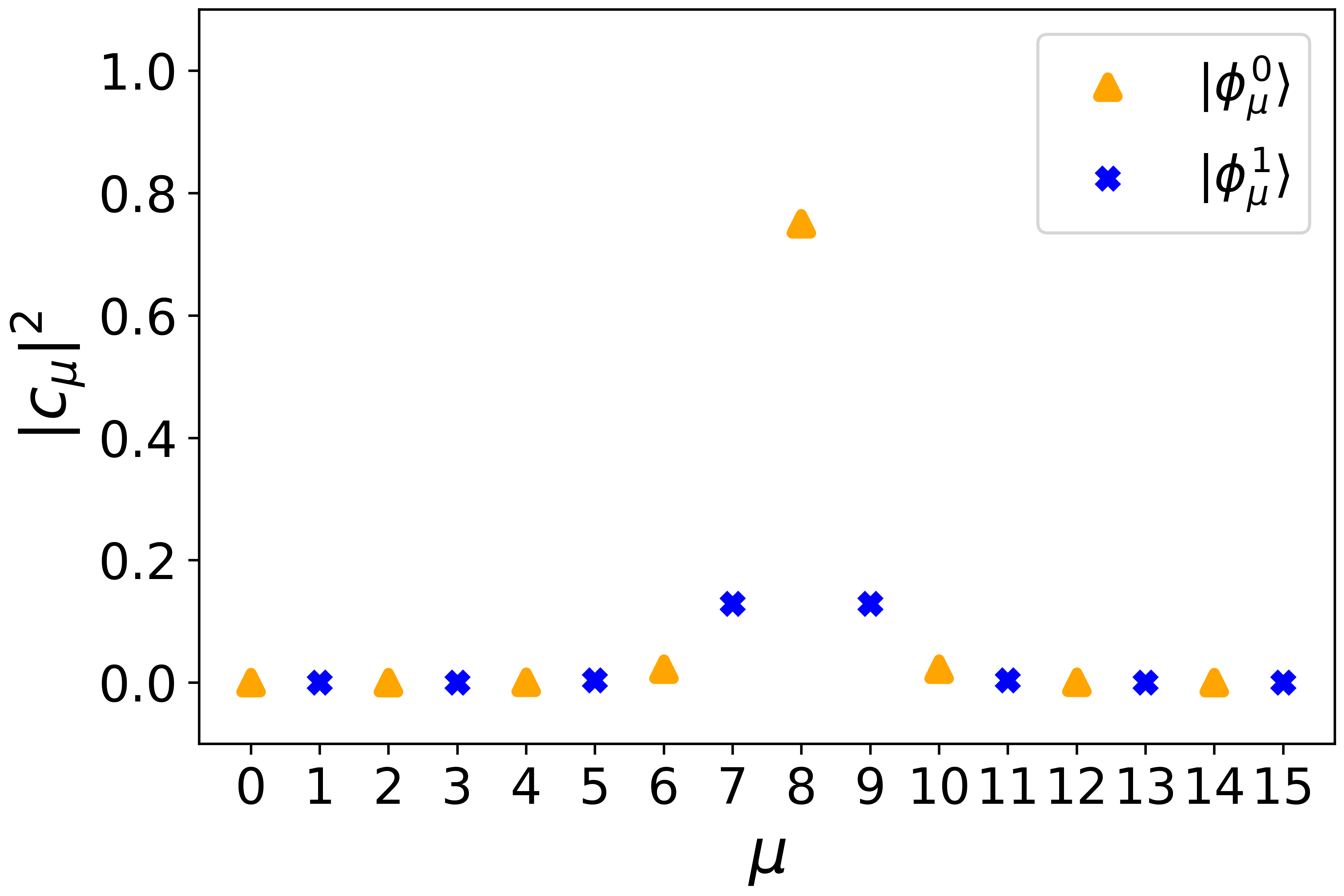}
\caption{Decomposition of a spin-flip state at a basal site into localized
magnons and non-stationary localized magnons according to \fmref{E-3-6}.
}
\label{decomposition}
\end{figure}

Figure~\xref{decomposition} demonstrates for an example 
of a spin-flip state at a basal site
how the coefficients
$c_\mu^{(j)}$ fall off with growing distance $|\mu-j|$ from the site of the spin flip. 
The overwhelming weight is indeed taken by the localized magnon 
at that position, i.e.\ $\mu=j$. The two nearest localized magnons,
$\mu=j\pm 2$,
also carry some non-neglibile, but already much smaller weight. These contributions will
also remain localized for all times. The numbers given in \figref{decomposition}
can be directly related to the long term averages given in \figref{dyn-late-times}.

The case of a spin flip excitation at an apical side behaves very similarly 
and is therefore not shown. As anticipated, the contributions of the two localized
magnons connected to that apical site is indeed largest, and contributions from 
localized magnons further away again fall off very rapidly.

\section{Discussion and conclusions}
\label{sec-4}

In this article, we demonstrated that certain carefully prepared Hamiltonians
show non-ergodic dynamics in contrast to the vast number of generic Hamiltonians
nearby in some parameter space. In our demonstration, the behavior can be traced 
back to the influence of a perfectly flat energy band that is characterized
by zero group velocity or equivalently by independent localized one-magnon states 
that are eigenstates of the Hamiltonian and therefore stationary. The latter
phenomenon has thus been termed ``disorder-free localization". It is an interference
effect due to the fine-tuned frustration of the competing interactions $J_1$ and $J_2$
\cite{DRM:IJMPB15}.

Although we only investigated the time-evolution of single-spin flips 
on the background of a magnon vacuum the results can be easily transferred
to arbitrary initial states since these can be written as superpositions 
of single-spin flip states.

Flat bands appear for all kinds of Hamiltonians and have initially been 
investigated for the Hubbard model \cite{MiT:CMP93}. It is therefore
no surprise that observations similar to ours 
have been discussed in connection with Hubbard models \cite{Dau:Dis22}.
Many flat-band systems have a realization as a magnetic material, for
instance kagome or pyrochlore systems. Recently, the idea was brought up
that Hamiltonians of such systems can be tuned by electric fields in 
order to set up a flat-band scenario \cite{ROS:PRB22}. This can potentially 
be achieved with multiferroic materials as e.g.\ discussed in \cite{MMB:PRM23}.

\section*{Acknowledgment}

This work was supported by the Deutsche Forschungsgemeinschaft DFG
(355031190 (FOR~2692); 397300368 (SCHN~615/25-2)).
We thank Masud Haque and Johannes Richter for discussions.


%

\end{document}